# Balance between noise fluxes in free-running single-mode class-A lasers

J. Jahanpanah, A. Soleimani, and F. Shavandi, *Member, IEEE*

*Abstract*— **The fluctuation effect of laser pumping rate on the output noise fluxes of class-A lasers is investigated. The method is based on the role of cavity Langevin force as a fluctuating force in the absence of the atomic population inversion and dipole moment Langevin forces. The temporal fluctuations induced to the phase and amplitude of the cavity electric field and the atomic population inversion are calculated in both below and above threshold states. Our aim is to derive correlation functions for the fluctuating variables of the cavity electric field and the atomic population inversion to determine the noise fluxes emerging from the cavity mirrors and measured by an optical detector and those radiated in the form of spontaneous emission in all spatial directions. We introduce a heuristic conservation relation that connects the noise flux generated by the laser pumping system with those distributed among the laser variables. Finally, the results are confirmed by demonstrating the energy conservation law.**

## I. INTRODUCTION

Damping is an unavoidable common problem of all quantum oscillators, which is always accompanied by fluctuation. The fluctuation is then converted into random changes (noise) in the phase and amplitude of periodic variables. The noise features of oscillators are especially much more complicated when the number of coupled oscillatory variables is increased [1], [2].

Lasers consist of three important variables of the cavity electric field $\alpha$ , the atomic population inversion $D$, and the atomic dipole moment $d$ with the respective damping rates $\gamma_C$, $\gamma_\|$, and $\gamma_\perp$ that are coupled to each other by the usual Maxwell-Bloch equations of motion [3], [4]. The origin of their fluctuations are, respectively, the well-known Langevin forces $\Gamma_\alpha$, $\Gamma_D$, and $\Gamma_d$ [5]. The correlation function of cavity Langevin force $\Gamma_\alpha$ is proportional to the mean number of thermal photons so that it is usually neglected at the operation temperature of lasers [1], [6]. Therefore, it was used to work with the atomic population inversion $\Gamma_D$ and dipole moment $\Gamma_d$ Langevin forces as the fluctuating forces [6]-[8].

On the other hand, the aim of this work is to demonstrate some interesting aspects of noise that can only be illustrated by considering the cavity Langevin force $\Gamma_\alpha$ alone. We thus remove the Langevin forces of $\Gamma_D$ and $\Gamma_d$ from the Bloch equations of motion to reveal the new aspects of laser noise with regard to the cavity Langevin force of $\Gamma_\alpha$. For simplicity, we also consider class-A lasers [9]-[11] in which the condition $\gamma_\perp \rangle\rangle \gamma_\| \rangle\rangle \gamma_c$ causes the Bloch equations of motion associated with the population inversion and the dipole moment of atoms are adiabatically approximated and substituted into the Maxwell-equation of motion pertaining to the cavity electric field.

There are many experiments to confirm that both the optical and electrical pumping systems generate the major noise of lasers. So, the different attempts have been made to reduce the laser noise produced by the pumping system [12]-[14]. It is due to the atomic population of inversion and the cavity electric field that are simultaneously controlled by the laser pumping rate. Any sudden change (fluctuation) in the

J. Jahanpanah. is with the Physics Department, Tarbiat Moallem University, 49 Mofateh Ave, Tehran, Iran (e-mail: jahanpanah@tmu.ac.ir).
A. Soleimani is a PhD student in Physics Department, Tarbiat Moallem University, 49 Mofateh Ave, Tehran, Iran (e-mail a. soleimani@tmu.ac.ir:)
F. Shavandi is a M.Sc student in Physics Department, Tarbiat Moallem University, 49 Mofateh Ave, Tehran, Iran (e-mail: f.shavandi@yahoo.com).

pumping variable is directly transferred into the two other laser variables. As a result, we are looking for a conservation rate between the input noise flux imposed to laser by fluctuations of the pumping system and the output noise fluxes distributed between the cavity electric field and the atomic population inversion in both below and above threshold states.

The paper is sorted out according to the following sections. In section II, the Maxwell-Bloch equations of motion for all the three class A, B and C lasers are introduced and then simplified to cover the motion of class-A lasers. The trial solutions for the fluctuating variables in the below threshold state are given in section III. The noise fluxes emerging through the cavity mirrors in the presence and the absence of laser pumping rate are also calculated. In section IV, we introduce the noise equations of motion and derive their solutions for the above threshold state. The conservation relation between the input and output energy rates of laser and their corresponding fluctuations are discussed in details in section V. Consequently, we will demonstrate a balance between the noise fluxes produced by the different fluctuating variables. Finally, the main results are summarized in section VI.

## II. THE MOTION EQUATIONS OF CLASS-A LASERS

The general features of the Maxwell-Bloch equations of motion including the fluctuating variables of the cavity electric field $\alpha$, the atomic population inversion $D$, and the atomic dipole moment $d$ together with their respective damping rates $\gamma_C$, $\gamma_{\|}$, and $\gamma_{\perp}$; and Langevin forces $\Gamma_\alpha$, $\Gamma_D$, and $\Gamma_d$ as the fluctuating forces are given by [15]-[18]

$$\dot{\alpha}+\left(\gamma_C+i\omega_L\right)\alpha=gd+\Gamma_\alpha, \tag{1}$$

$$\dot{D}+\gamma_{\|}D=\gamma_{\|}D_P-g\left(\alpha^* d+\alpha d^*\right)+\Gamma_D, \tag{2}$$

and

$$\dot{d}+\left(\gamma_{\perp}+i\omega_L\right)d=g\alpha D+\Gamma_d, \tag{3}$$

in which $\omega_L$ is the laser oscillation frequency, $D_P$ is the inverted atomic population produced by the laser pumping system and $g$ is the coupling constant between the cavity electric field and the atomic dipole moment.

The real role of Langevin force $\Gamma_\alpha$ is revealed by ignoring two other Langevin forces ($\Gamma_D = \Gamma_d = 0$). We also apply the condition $\gamma_\perp \rangle\rangle \gamma_{\|} \rangle\rangle \gamma_c$ associated with class-A lasers to (3). Then, the dipole moment variable is immediately derived as [3], [9]

$$d=g\alpha D/\gamma_{\perp}. \tag{4}$$

By substituting $d$ from (4) into (1) and (2), we will have

$$\dot{\alpha}+\left(\gamma_C+i\omega_L\right)\alpha=\frac{g^2}{\gamma_\perp}\alpha D+\Gamma_\alpha \tag{5}$$

and

$$\gamma_{\|}D=\gamma_{\|}D_P-\frac{2g^2}{\gamma_\perp}|\alpha|^2 D, \tag{6}$$

where $D$ is not directly substituted from (6) into (5) to avoid the nonlinear cubic term $|\alpha_L|^3$ that complicates the Fourier transform of (5) in subsequent sections.

## III. THE BELOW-THRESHOLD NOISE

The chaotic nature of laser light in below threshold state is distinguished by a zero-mean electric field that oscillates with a time-random amplitude $\delta\alpha(t)$ in the form

$$\alpha(t)=\delta\alpha(t)\exp(-i\omega_L t). \tag{7}$$

The population inversion gains a real time-dependent fluctuating term $\delta D(t)$ in the form

$$D(t)=D_P+\delta D(t). \tag{8}$$

If the trial solutions (7) and (8) are substituted into (5) and (6), then the terms proportional to first order of the fluctuating variables give

$$\delta\dot{\alpha}+\gamma_C(1-C)\delta\alpha(t)=\Gamma_\alpha(t)\exp(i\omega_L t) \tag{9}$$

and

$$\delta D(t)=0, \tag{10}$$

where $C=D_P/D_0$ ($D_0=\gamma_\perp\gamma_C/g^2$) is the normalized pumping rate and has a value less than one $(C<1)$ in the below and larger than one $(C\rangle 1)$ in the above threshold states [19].

The time-dependent variable $\delta\alpha(t)$ is related to its corresponding variable in frequency domain $\delta\alpha(\omega)$ by using the Fourier transform in the following formal form

$$\delta\alpha(\omega)=\frac{1}{\sqrt{2\pi}}\int dt\ \delta\alpha(t)\ \exp(i\omega t), \tag{11}$$

where $\omega$ is the frequency detuning with respect to the laser frequency $\omega_L$. It is evident from (10) and (11) that

$$\delta D(\omega)=0, \tag{12}$$

while (9) has a simple solution in the form

$$\delta\alpha(\omega)=\frac{\Gamma_\alpha(\omega_L+\omega)}{\gamma_C(1-C)-i\omega}. \tag{13}$$

The power spectrum of the cavity electric field is now obtained by the following correlation function

$$\left\langle\delta\alpha(\omega)\,\delta\alpha^*(\omega')\right\rangle=\frac{\left\langle\Gamma_\alpha(\omega_L+\omega)\Gamma_\alpha^{\ *}(\omega_L+\omega')\right\rangle}{[\gamma_C(1-C)-i\omega][\gamma_C(1-C)+i\omega']}, \tag{14}$$

in which the correlation function of the cavity Langevin force $\Gamma_\alpha$ in nominator is given by Scully [1] and Louisell [6] as

$$\begin{aligned}\left\langle\Gamma_\alpha(\omega_L+\omega)\Gamma_\alpha^{\ *}(\omega_L+\omega')\right\rangle&=2\left\langle D_{\alpha\alpha^*}\right\rangle_R\delta(\omega-\omega')\\&=2\gamma_C(\bar{n}_{th}+1)\,\delta(\omega-\omega')\\&\approx 2\gamma_C\,\delta(\omega-\omega'),\end{aligned} \tag{15}$$

where the mean number of thermal photons $\bar{n}_{th}$ is usually ignored at the laser operation temperature according to $\bar{n}_{th}=1/\left[\exp\left(\hbar\omega_L/k_BT\right)-1\right]<<1$.

The correlation function of an arbitrary time-fluctuating variable $a(t)$ with a white noise origin (Dirac function) is defined in the following complex conjugate form [20]

$$\left\langle a^*(\omega)a(\omega')\right\rangle=2\pi h(\omega)h^*(\omega')\delta(\omega-\omega'), \tag{16}$$

in which $|h(\omega)|^2$ is the dimensionless mean flux per unit angular frequency bandwidth at angular frequency $\omega$ and given by

$$|h(\omega)|^2=\frac{1}{2\pi}\int d\omega'\left\langle a^*(\omega)a(\omega')\right\rangle\exp[i(\omega-\omega')t]. \tag{17}$$

Now by substituting (15) into (14), we derive the following correlation function similar to (16) as

$$\begin{aligned}\left\langle\delta\alpha(\omega)\,\delta\alpha^*(\omega')\right\rangle&=2\pi\,\frac{(\gamma_C/\pi)^{1/2}}{[\gamma_C(1-C)-i\omega]}\\&\times\frac{(\gamma_C/\pi)^{1/2}}{[\gamma_C(1-C)+i\omega']}\,\delta(\omega-\omega').\end{aligned} \tag{18}$$

Therefore, the output noise flux of class-A lasers in the below threshold state is given by

$$N_{LN}(\omega)=2\gamma_C\,|h(\omega)|^2=\frac{2\gamma_C^2}{\pi\left[\omega^2+\gamma_C^2(1-C)^2\right]}, \tag{19}$$

where the subscript $LN$ is an abbreviation for the laser noise. It is produced by the pumping noise flux $N_{PN}(\omega)$ (pumping noise) and can be measured by the optical detectors after passing through the cavity mirrors. The other part of pumping noise flux is converted into the noise flux of spontaneous emission $N_{SPN}(\omega)$ (spontaneous emission noise) and radiated into all spatial directions. This subject will be discussed in more details in next section.

The contribution of the pumping noise $N_{PN}(\omega)$ on the laser noise $N_{LN}(\omega)$ is now disappeared when the normalized pumping rate $C$ is approaching to zero, so that we have

$$N_{CN}(\omega)=\lim_{C\to 0}N_{LN}(\omega)=\frac{2\gamma_C^2}{\pi\left(\omega^2+\gamma_C^2\right)}, \tag{20}$$

which is in complete agreement with the Lorentzian spectrum of the empty cavity noise (cavity noise) $N_{CN}(\omega)$ derived by Scully and Zubairy in (9.3.12) of Ref [1]. As a result, a photodetector cannot distinguish the cavity noise $N_{CN}(\omega)$ and the laser noise $N_{LN}(\omega)$, independently. The laser noise spectra in the below threshold state for the different values of the normalized pumping rates $C$ ($C<1$)) are plotted in Fig. 1.

A similar relation compared to (19) was derived by Loudon and his co-workers (LHSV) by ignoring the cavity Langevin force $\Gamma_\alpha$ ($\Gamma_\alpha=0$, $\Gamma_D\neq 0$, and $\Gamma_d\neq 0$) [8]. It has only an additional multiplying factor of the normalized pumping rate $C$ on the nominator (19), as seen in (4.30) of Ref [8]. Consequently, the cavity noise (20) cannot be derived from the laser noise (19) in the absence of the normalized pumping rate ($C=0$ ).

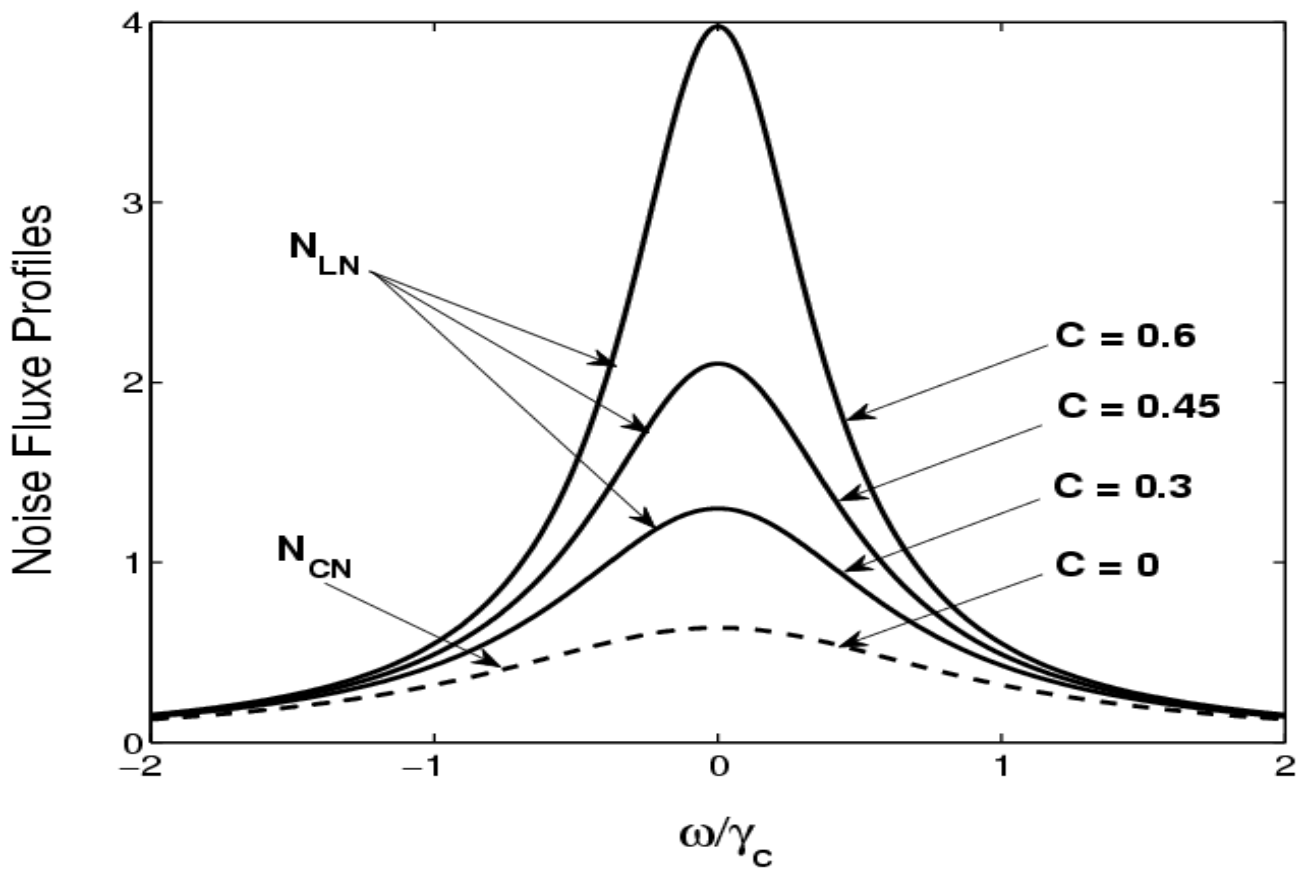

Fig. 1. The below-threshold profiles of noise fluxes against the normalized frequency detuning $\omega/\gamma_C$ are plotted in the absence of the laser pumping for the empty cavity with $C=0$ as shown by the dashed curve, and in the presence of laser pumping with $C$ = 0.3, 0.45, and 0.6 as shown by the solid curves.

Nevertheless, the bandwidth of laser noise spectrum is obtained from (19) as

$$\Delta\omega_{LN} = 2\gamma_C(1-C)\,, \tag{21}$$

which is in complete agreement with the similar relation (4.28) derived by LHSV [8]. As one expects, when $C$ tends to zero, the laser noise bandwidth $\Delta\omega_{LN}$ matches to the cavity noise bandwidth $\Delta\omega_{CN} = 2\gamma_C$.

According to (15), it should be noted that the laser noise flux (19) is only created by the mean damping rate of the cavity mirrors $\gamma_C$, rather by the negligible mean number of the thermal photons $\bar{n}_{th}$. Therefore, we can not ignore from the vital role of the cavity langevin force $\Gamma_\alpha$ on the laser noise.

## IV. THE ABOVE-THRESHOLD NOISE

The above threshold state of laser is specified by a strong coherent light of the mean large amplitude $|\alpha_L|$ produced by the stimulated emission radiation. However, it also consists of two real fluctuating variables of amplitude $\delta\alpha(t)$ and phase $\delta\phi(t)$ produced by the laser noise, so that the cavity electric field is defined in the following complete form

$$\alpha(t) = [|\alpha_L| + \delta\alpha(t)]\exp[-i\omega_L t + i\delta\phi(t)]\,, \tag{22}$$

where $|\alpha_L|^2$ is the mean number of the cavity photons in the above threshold state ($C>1$) and given by [11] and [16] as

$$|\alpha_L|^2 = \frac{\gamma_\perp\gamma_\parallel}{2g^2}(C-1)\,. \tag{23}$$

The static population inversion of atoms $D_O$ is also fluctuated by a real time-dependent random value $\delta D(t)$ as [11], [16]

$$D(t) = D_0 + \delta D(t)\,. \tag{24}$$

If one substitutes the trial solutions (22) and (24) into (5) and (6), the following equations proportional to the first order in fluctuating variables $\delta\alpha(t), \delta\varphi(t)$ and $\delta D(t)$ are obtained in the forms

$$\begin{aligned}\delta\dot{\alpha}(t) + i|\alpha_L|\delta\dot{\varphi}(t) = \frac{g^2}{\gamma_\perp}|\alpha_L|\,\delta D(t) \\ + \Gamma_\alpha(t)\exp[i\omega_L t - i\delta\phi(t)]\end{aligned} \tag{25}$$

and

$$\gamma_\parallel C\,\delta D(t) = -4\gamma_C|\alpha_L|\,\delta\alpha(t)\,. \tag{26}$$

The variables $\delta\alpha(\omega)$ and $\delta\phi(\omega)$ are now derived from the real and imaginary parts of the Fourier transform (25) as

$$\delta\alpha(\omega) = \frac{1}{2[2\gamma_C(C-1)/C - i\omega]}\left[\Gamma_\alpha(\omega_L+\omega) + \Gamma_\alpha^*(\omega_L+\omega)\right] \tag{27}$$

and

$$\delta\phi(\omega) = \frac{1}{2|\alpha_L|\omega}\left[\Gamma_\alpha(\omega_L+\omega) - \Gamma_\alpha^*(\omega_L+\omega)\right]. \tag{28}$$

The variable $\delta D(\omega)$ is then calculated by substituting $\delta\alpha(\omega)$ from (27) into the Fourier transform of (26) as

$$\delta D(\omega) = \frac{-2\gamma_C|\alpha_L|}{\gamma_\parallel C[2\gamma_C(C-1)/C - i\omega]}\left[\Gamma_\alpha(\omega_L+\omega) + \Gamma_\alpha^*(\omega_L+\omega)\right]\cdot \tag{29}$$

Now by using (15) and the following correlation function that is defined in Refs of [1], [6], and [7] as

$$\begin{aligned}\left\langle\Gamma_\alpha^{\,*}(\omega_L+\omega)\,\Gamma_\alpha(\omega_L+\omega')\right\rangle &= 2\left\langle D_{\alpha^*\alpha}\right\rangle_R\delta(\omega-\omega')\,, \\ &= 2\gamma_C\,\bar{n}_{th}\,\delta(\omega-\omega')\end{aligned} \tag{30}$$

we derive the correlation functions for the variables of $\delta\alpha(\omega)$, $\delta\phi(\omega)$, and $\gamma_\parallel\delta D(\omega)$ in the following respective forms

$$\left\langle \delta\alpha^*(\omega)\,\delta\alpha(\omega')\right\rangle = 2\pi \frac{(\gamma_C/\pi)^{1/2}}{2[2\gamma_C\,(C-1)/C + i\omega]} \times \frac{(\gamma_C/\pi)^{1/2}}{2[2\gamma_C\,C-1/C - i\omega']}\,\delta(\omega-\omega'), = 2\pi\, h_{LN}(\omega)\, h^*_{LN}(\omega')\delta(\omega-\omega') \quad (31)$$

$$\left\langle \delta\phi^*(\omega)\,\delta\phi(\omega')\right\rangle = \frac{\gamma_C}{|\alpha_L|^2\,\omega^2}\,\delta(\omega-\omega'), \quad (32)$$

and

$$\frac{\left\langle [\gamma_\parallel \delta D^*(\omega)][\gamma_\parallel \delta D(\omega')]\right\rangle}{2\gamma_C|\alpha_L|^2} = 2\pi \frac{(2\gamma_C^2/\pi)^{1/2}}{C[2\gamma_C\,(C-1)/C + i\omega]} \times \frac{(2\gamma_C^2/\pi)^{1/2}}{C[2\gamma_C\,C-1/C - i\omega']}\,\delta(\omega-\omega') = 2\pi\, h_{SPN}(\omega)\, h^*_{SPN}(\omega')\,\delta(\omega-\omega') \quad (33)$$

where the condition $\overline{n_{th}} \langle\langle 1$ is used. The phase correlation function of (32) is the same as (5.50) derived by LHSV [8]. Therefore, we here give up from the fluctuation effect of the electric field phase (known as the phase diffusion effect or the Schawlow-Townes linewidth), since it had already discussed by many authors, especially in [8], [21], and [22].

The output noise flux from the cavity mirrors due to the fluctuations of the electric field amplitude $\delta\alpha(t)$ is given by (31) as

$$N_{LN}(\omega) = 2\gamma_C\,|\,h_{LN}(\omega)\,|^2 = \frac{\gamma_C^2}{2\pi\left[\omega^2 + 4\gamma_C^2\left(\frac{C-1}{C}\right)^2\right]}. \quad (34)$$

The spontaneous emission noise flux radiated in all spatial directions due to the fluctuations in the static population inversion of atoms $\delta D(\omega)$ is given by (33) as

$$N_{SPN}(\omega) = |\,h_{SPN}(\omega)\,|^2 = \frac{2\gamma_C^2}{\pi C^2\left[\omega^2 + 4\gamma_C^2\left(\frac{C-1}{C}\right)^2\right]}. \quad (35)$$

The noise fluxes of $N_{LN}(\omega)$ and $N_{SPN}(\omega)$ versus the normalized frequency detuning $\omega/\gamma_C$ for the different values of the normalized pumping rate $C$ are plotted in Figs. 2(a) and 2(b), respectively.

## V. THE FLUCTUATION AND NOISE CONSERVATION RELATIONS

The energy conservation law requires that the total energy rate supplied into laser by the pumping system $\gamma_\parallel D'_P(t)$ is equal to the total energy rate emitted from the laser in the forms of spontaneous emission rate $\gamma_\parallel D(t)$ in all spatial directions and the stimulated emission rate $2\gamma_C|\alpha(t)|^2$ emerged through the cavity mirrors, so that we have [3], [11]

$$\gamma_\parallel D'_P(t) = \gamma_\parallel D(t) + 2\gamma_C|\alpha(t)|^2, \quad (36)$$

in which the pumping variable $D'_P(t)$ includes the time-independent mean value $D_P$ together with a fluctuating value $\delta D_P(t)$ as

$$D'_P(t) = D_P + \delta D_P(t), \quad (37)$$

where it holds for the both below and above threshold states.

In below threshold state, the fluctuation conservation relation is derived by substituting the corresponding trial solutions (7), (8), and (37) into the Fourier transform of (36) as

$$\gamma_\parallel \delta D_P(\omega) = \gamma_\parallel \delta D(\omega) + 2\gamma_C \delta\alpha^2(\omega) \approx \gamma_\parallel \delta D(\omega)\,. \quad (38)$$

Therefore, correct to first order of fluctuating variables, any fluctuation in the laser pumping rate $\gamma_\parallel \delta D_P(\omega)$ directly transfers to the spontaneous emission rate $\gamma_\parallel \delta D(\omega)$.

In above threshold state, the fluctuation conservation relation correct to the first-order fluctuating variables of the pumping rate $\gamma_\parallel \delta D_P(t)$, the spontaneous emission rate $\gamma_\parallel \delta D(t)$, and the electric field flux $4\gamma_C|\alpha_L|\delta\alpha(t)$ is directly derived by substituting the trial solutions (22), (24), and (37) into the energy conservation relation (36) in the form

$$\gamma_\parallel \delta D_P(t) = \gamma_\parallel \delta D(t) + 4\gamma_C\,|\,\alpha_L\,|\,\delta\alpha(t)\,, \quad (39)$$

or in the equivalent form

$$\gamma_\parallel \delta D_P(\omega) = \gamma_\parallel \delta D(\omega) + 4\gamma_C\,|\,\alpha_L\,|\,\delta\alpha(\omega)\,. \quad (40)$$

It is evident from (22) and (36) that why any contribution of the fluctuating variable $\delta\phi(t)$ has not appeared in the energy and fluctuation conservation relations (36) and (40),

(a)

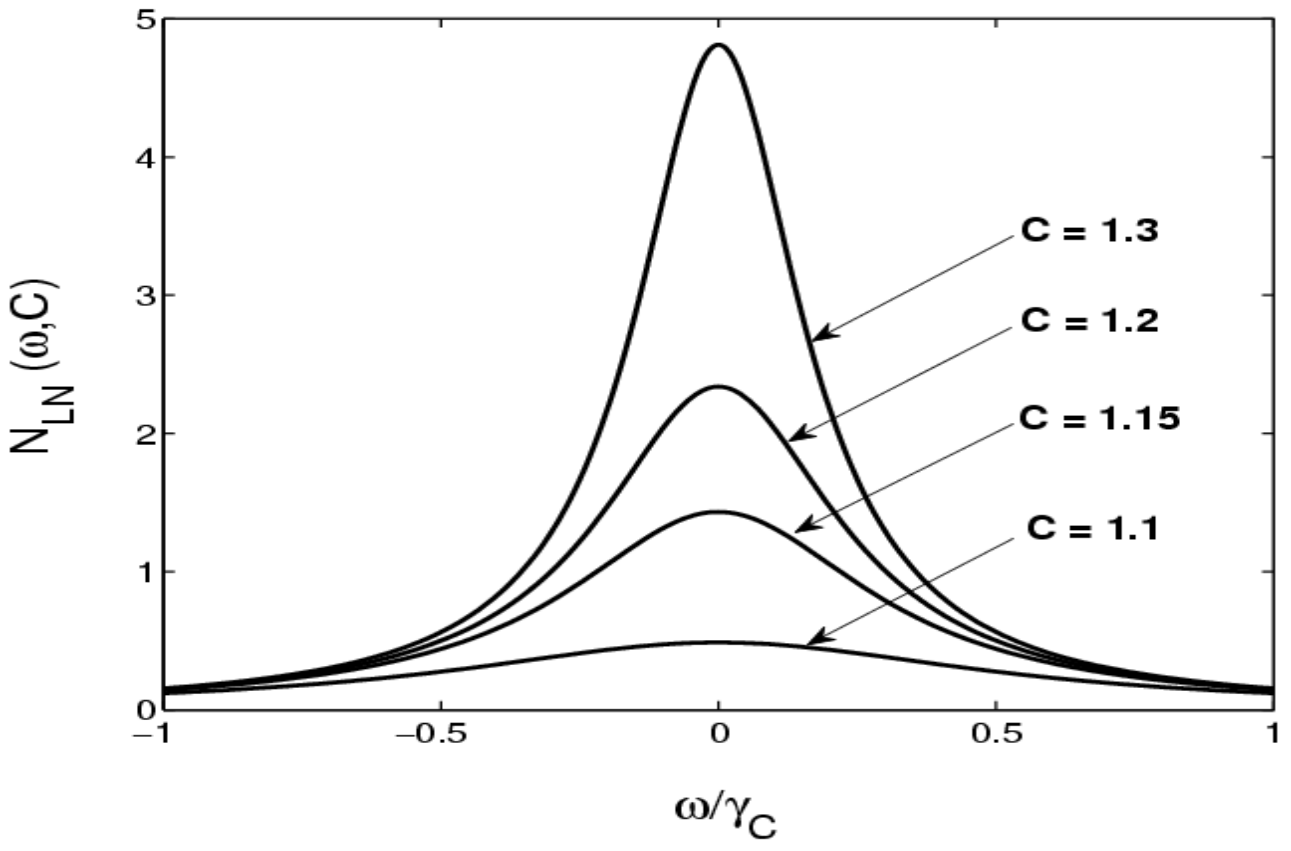

(b)

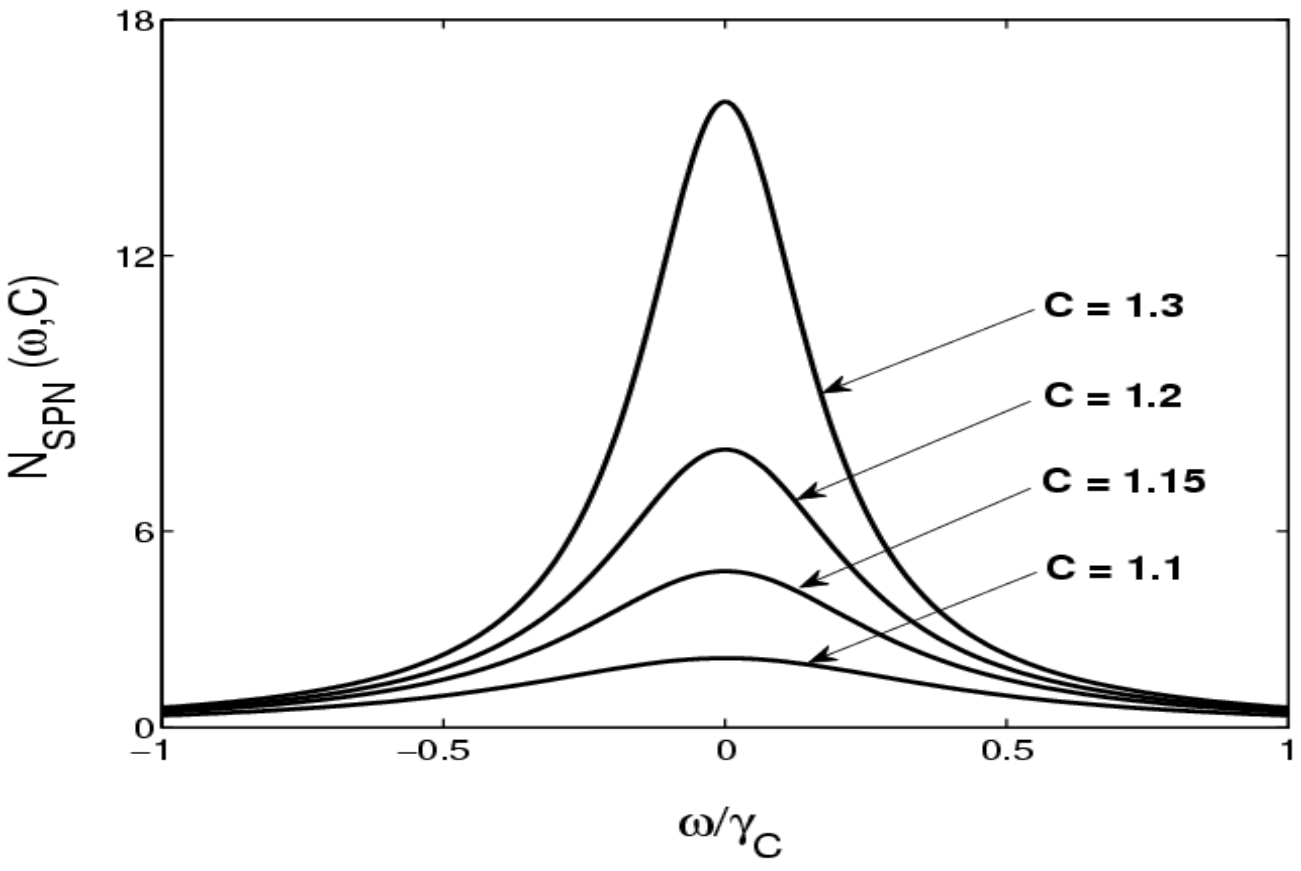

respectively. The effect of the phase fluctuations is to offset the laser emission frequency (see, for example, page 294 of Ref [23]).

The fluctuation rate of the laser pumping is now determined by substituting $\delta\alpha(\omega)$ and $\delta D(\omega)$ from (27) and (29) into the fluctuation conservation relation (40) as

$$\gamma_{\|}\delta D_P(\omega)=\frac{2\gamma_C|\alpha_L|(C-1)/C}{\left[-i\omega+2\gamma_C\,\frac{C-1}{C}\right]}\left[\Gamma_\alpha(\omega_L+\omega)+\Gamma_\alpha^*(\omega_L+\omega)\right]. \tag{41}$$

We then derive the following correlation function for the laser pumping fluctuation by using the correlation relations (15) and (30) in the form

(c)

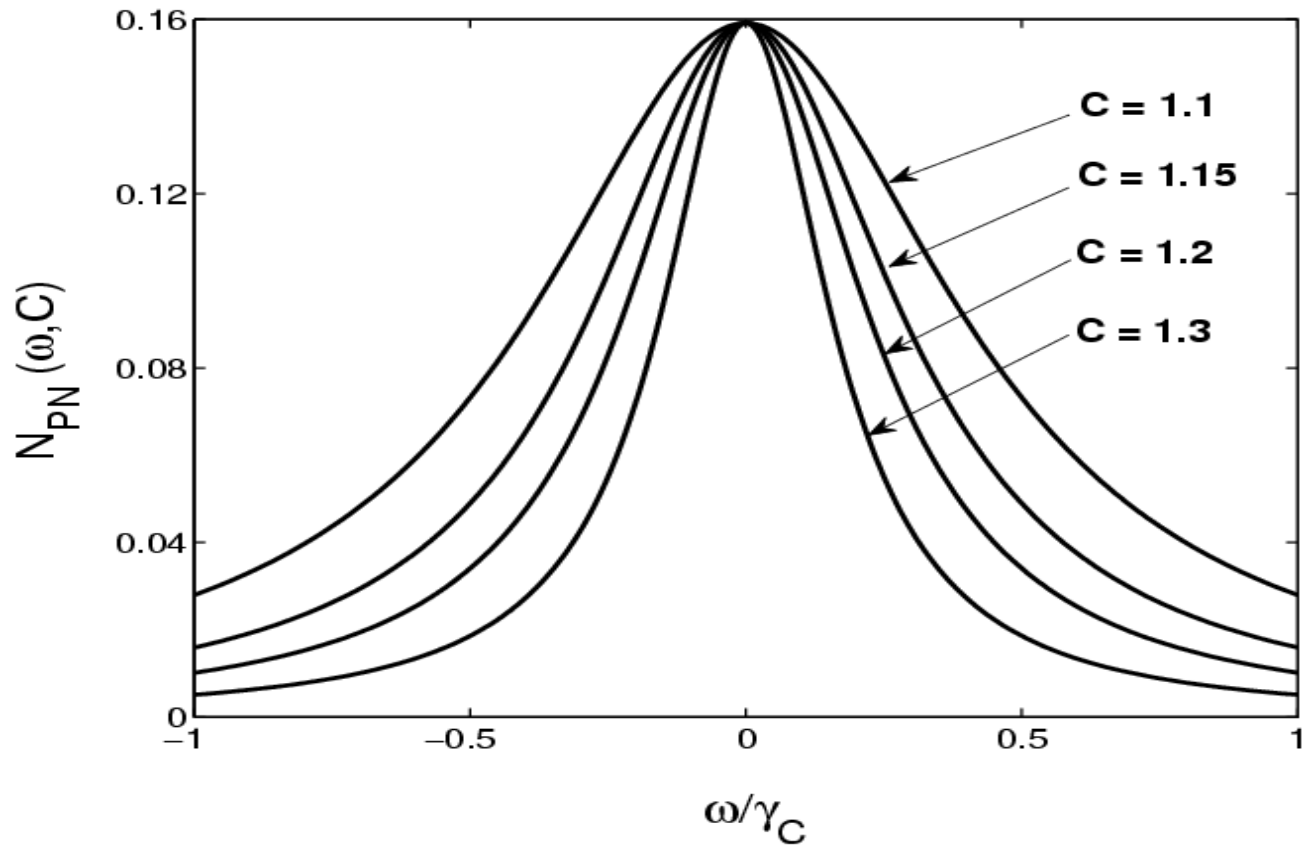

Fig. 2. The above-threshold profiles of noise fluxes against the normalized frequency detuning $\omega/\gamma_C$ are illustrated for the normalized pumping rates $C$ = 1.1, 1.15, 1.2, and 1.3 for (a). the laser noise $N_{LN}(\omega/\gamma_C)$, (b). the spontaneous emission noise $N_{SPN}(\omega/\gamma_C)$, and (c) the pumping noise $N_{PN}(\omega/\gamma_C)$.

$$\begin{aligned}\frac{\left\langle[\gamma_{\|}\delta D_P{}^*(\omega)][\gamma_{\|}\delta D_P(\omega')]\right\rangle}{2\gamma_C|\alpha_L|^2}&=2\pi\frac{\left[2\gamma_C^2(C-1)^2/(C^2\pi)\right]^{1/2}}{\left[i\omega+2\gamma_C\,(C-1)/C\right]}\\&\quad\times\frac{\left[2\gamma_C^2(C-1)^2/(C^2\pi)\right]^{1/2}}{\left[-i\omega'+2\gamma_C\,(C-1)/C\right]}\delta(\omega-\omega')\\&=2\pi\,h_{PN}(\omega)\,h_{PN}^*(\omega')\,\delta(\omega-\omega').\end{aligned} \tag{42}$$

The spectrum of pumping noise flux is thus given by

$$N_{PN}(\omega)=|\,h_{PN}(\omega)\,|^2=\frac{2\gamma_C^2\,(C-1)^2}{\pi C^2\left[\omega^2+4\gamma_C^2\left(\frac{C-1}{C}\right)^2\right]}. \tag{43}$$

The variations of $N_{PN}(\omega)$ versus $\omega/\gamma_C$ for the different values $C$ are demonstrated in Fig. 2(c).

The next aim is to derive a conservation relation between the noise fluxes $N_{PN}(\omega)$, $N_{LN}(\omega)$, and $N_{SPN}(\omega)$ by taking the complex conjugate of fluctuation conservation relation (40) at an arbitrary frequency $\omega'$ as

$$\gamma_{\|}\,\delta D_P^*(\omega')=\gamma_{\|}\,\delta D^*(\omega')+4\,\gamma_C\,|\,\alpha_L\,|\,\delta\alpha^*(\omega'). \tag{44}$$

Now by multiplying the fluctuation conservation relations (40) and (44) in each other and taking the average of its complex conjugate, we derive the following relation

$$\left\langle [\gamma_{\parallel}\delta D_P{}^{*}(\omega)][\gamma_{\parallel}\delta D_P(\omega')]\right\rangle = \left\langle [\gamma_{\parallel}\delta D^{*}(\omega)][\gamma_{\parallel}\delta D(\omega')]\right\rangle + 16\gamma_C^2\,|\alpha_L|^2\left\langle \delta\alpha^{*}(\omega)\,\delta\alpha(\omega')\right\rangle + 4\gamma_C|\alpha_L|\left[\left\langle[\gamma_{\parallel}\delta D^{*}(\omega)]\delta\alpha(\omega')\right\rangle + \left\langle \delta\alpha^{*}(\omega)[\gamma_{\parallel}\delta D(\omega')]\right\rangle\right]. \tag{45}$$

Finally, if the above relation is divided to the mean output flux of photons $2\gamma_C|\alpha_L|^2$ and relations (34), (35), and (43) are used, after some algebra calculations, we derive a heuristic relation for the conservation of the noise fluxes as

$$N_{PN}(\omega) = N_{SPN}(\omega) + 4\left(\frac{C-2}{C}\right)N_{LN}(\omega)\cdot \tag{46}$$

It describes how the input noise flux of laser pumping $N_{PN}(\omega)$ shares between the noise flux of the spontaneous emission $N_{SPN}(\omega)$ radiated in all spatial directions and the noise flux of the cavity electric field $N_{LN}(\omega)$ passed through the cavity mirrors and measured by a photodetector. It is easy to confirm that the noise flux conservation relation (46) even holds at the laser threshold state with the normalized pumping rate $C=1$, where the spectrum of pumping noise flux is cancelled out ($N_{PN}(\omega)=0$) by sum of the two other noise flux spectra $N_{SPN}(\omega) = 2\gamma_C^2/\pi\omega^2$ and $N_{LN}(\omega) = \gamma_C^2/2\pi\omega^2$ determined by (34) and (35) at $C=1$.

Fig. 3 demonstrates the variations of the laser noise $N_{LN}(\omega)$, the spontaneous emission noise $N_{SPN}(\omega)$, and the pumping noise $N_{PN}(\omega)$ versus the normalized detuning frequency $\omega/\gamma_C$ for $C=1.3$. It is evident that they satisfy the conservation relation of the noise flux (46). However, the bandwidths of all three noise fluxes $N_{PN}(\omega)$, $N_{SPN}(\omega)$, and $N_{LN}(\omega)$ are equal and given by

$$(\Delta\omega)_{PN} = (\Delta\omega)_{SPN} = (\Delta\omega)_{LN} = 4\gamma_C\,\frac{C-1}{C}, \tag{47}$$

which is in agreement with the laser noise bandwidth (5.62) derived by LHSV [8]. The bandwidth variations of the noise fluxes (21) and (47) in the below and above threshold states are plotted against the normalized pumping rate $C$ in Fig. 4. It is seen that all bandwidths reduce to zero at the laser threshold state $C=1$, and increase by moving away from threshold ($C=1$) toward smaller ($C<1$) and larger ($C>1$) values of the normalized pumping rate $C$. The maximum values of the bandwidths are equal to $2\gamma_C$ at the below-threshold state for $C=0$ and equal to $4\gamma_C$ at the above-threshold state for $C>>1$.

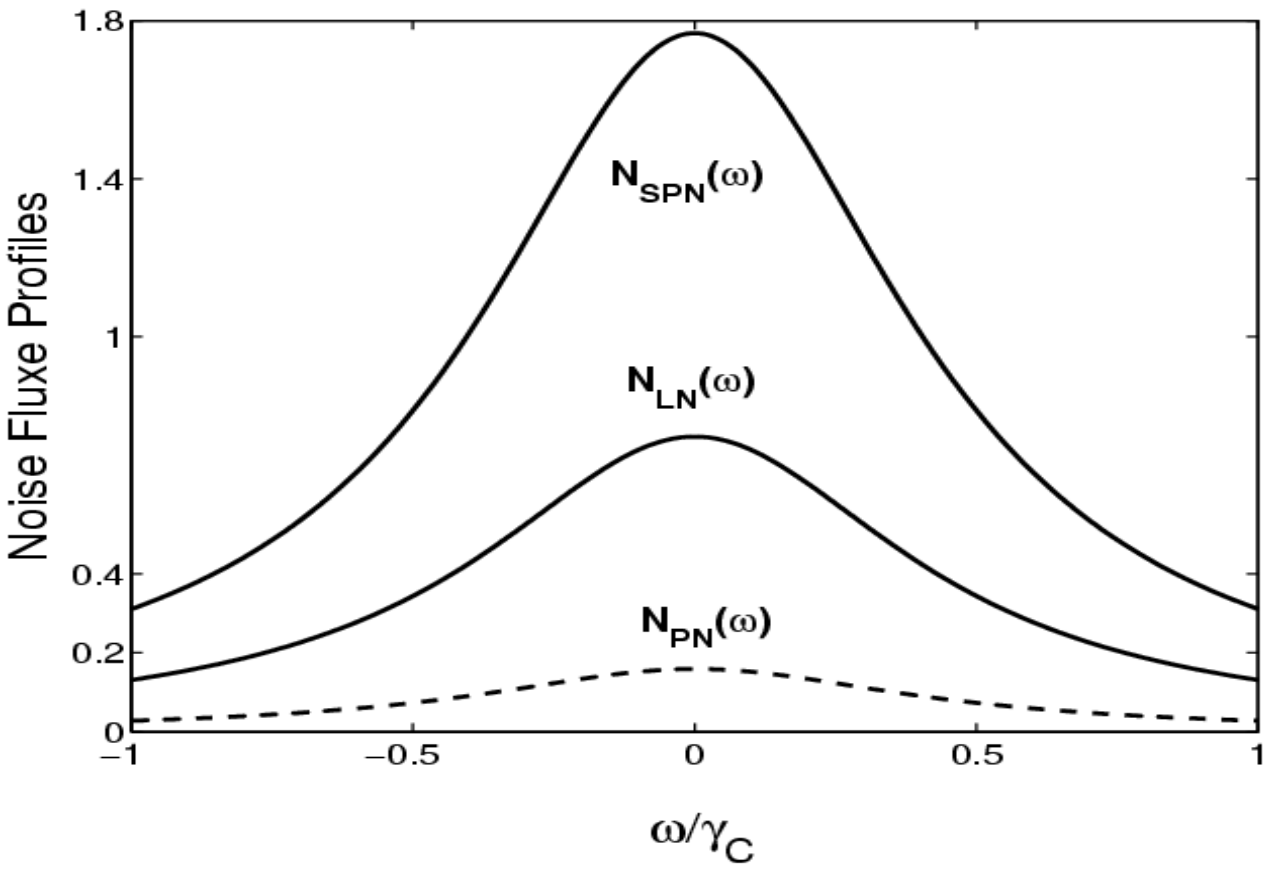

Fig. 3. The pumping noise flux $N_{PN}(\omega/\gamma_C)$ is calculated as sum of $N_{LN}(\omega/\gamma_C)$ and $N_{SPN}(\omega/\gamma_C)$ by using the noise flux conservation relation (46) and the noise definitions (34) and (35) for $C=1.3$. The pumping noise profile (dashed curve) is in complete agreement with its corresponding curve plotted in the part (c) of Fig (2) for $C=1.3$.

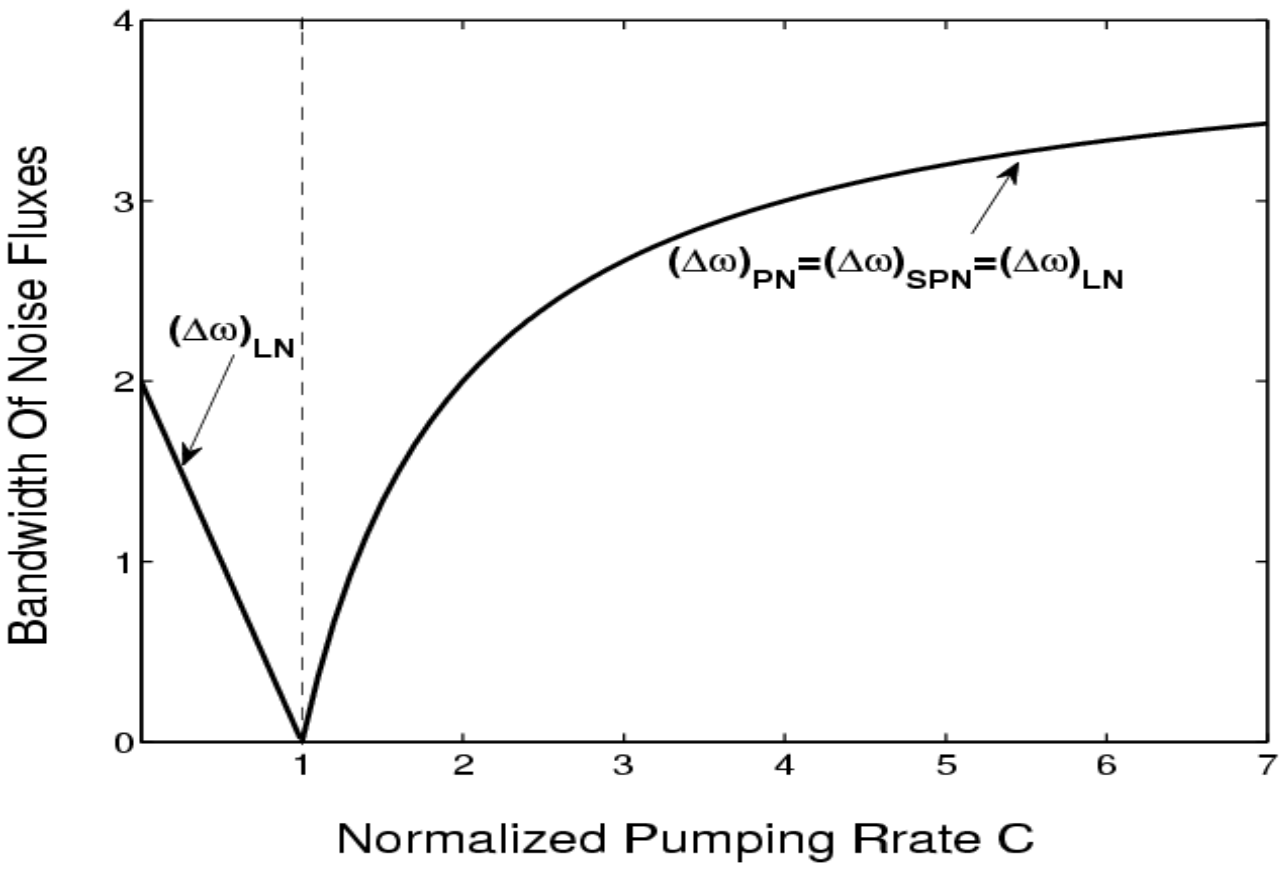

Fig. 4. The bandwidth of noise fluxes versus the normalized pumping rate $C$ is shown for (a) the below-threshold state with $C<1$, and for (b) the above-threshold state with $C>1$.

## VI. CONCLUSION

We have introduced a new method which describes how the noise flux produced by the laser pumping system is distributed between the spontaneous and stimulated emission radiations. The noise produced by the excitations of the thermal photons ($n_{th}$) is always ignorable, but the noise produced by the mean decay rate of the cavity mirrors ($\gamma_C$) plays the significant role. This is due to the definition of cavity diffusion coefficient $\left\langle D_{\alpha\alpha^*} \right\rangle = \gamma_C (n_{th} + 1) \approx \gamma_C$ in which $n_{th} << 1$ whereas $\gamma_C \approx 10^6 - 10^8\, Hz$. Therefore, we have only considered the cavity Langevin force $\Gamma_\alpha$ ($\Gamma_D = \Gamma_d = 0$) as a fluctuating force for the variables of the laser pumping, the atomic population inversion, and the cavity electric field in the both below and above threshold states of a class-A laser. The results are in good agreement with the case of $\Gamma_D \neq 0$, $\Gamma_d \neq 0$, and $\Gamma_\alpha = 0$ [8].

Finally, we have demonstrated that the noise fluxes of pumping $N_{PN}(\omega)$, laser $N_{LN}(\omega)$, and spontaneous emission $N_{SPN}(\omega)$ satisfy the flux conservation relation (46) for all values of the normalized pumping rates $C$.

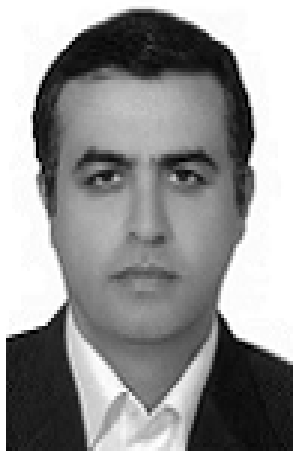

**Jafar Jahanpanah** received the B.sc. degree in Applied physics from Amir Kabir (Tehran Poly Technics) University in 1989, the M.sc. degree in Optics communication in 1992, and the PhD degree in Laser theory in 1995, both from Essex University in UK.

He has achieved as a senior student to gain a scholarship from the Iranian culture and higher education minister to continue his education career at Essex University in 1991. He has begun his PhD research with Prof. Rodney Loudon on the gain, stability, and injection-locking theory of single-mode laser amplifiers from 1992 to 1995. His research is now extended to cover the laser noise and also the gain, stability, and mode-locking phenomena in three-mode lasers with the publications in Journals of PRA, OSA, Applied Physics, and Optics Communications.

Dr. Jahanpanah has been Assistant Professor of the Physics Department at Tarbiat Moallem University (TMU) in 1997 and appointed as the research chairman of the Science Faculty at TMU from 2006 to 2009. He is now Associate Professor of physics and the chairman of the Science Faculty at TMU since 2009.

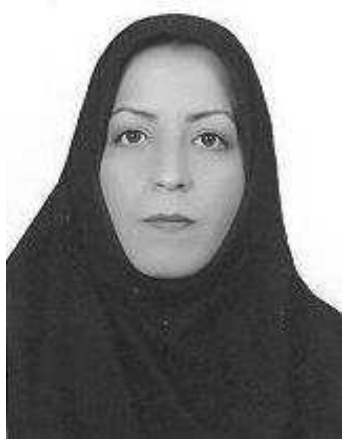

**Azam soleimani** have the B.Sc. degree in physics from Tarbiat Moallem University (2002) and the M.Sc. degree in atomic and molecular physics from Iran University of Science and Technology(2004). Now, she is a PhD student in Physics Department of Tarbiat Moallem University.

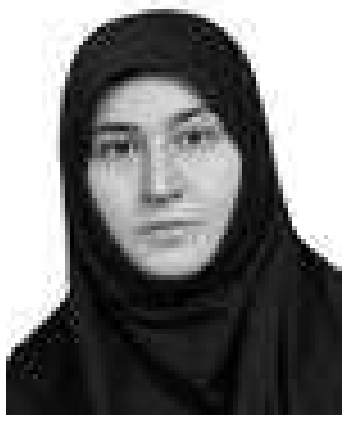

**Fatemeh Shavandi** have the B.Sc. degree in physics from Azad University (2009) and now she is a M.Sc student in Physics Department of Tarbiat Moallem University.